\documentclass{article}
\usepackage{ismir,amsmath,cite,url}
\usepackage{graphicx}
\usepackage{color}
\usepackage{amssymb}
\usepackage{stmaryrd}

\usepackage{xspace}
\usepackage{bbm}
\def\eg{e.g.\@\xspace}
\def\ie{i.e.\@\xspace}
\def\resp{resp.\@\xspace}
\def\vs{vs.\@\xspace}

\title{Eigentriads and Eigenprogressions on the Tonnetz}

\oneauthor
 {Vincent Lostanlen}
 {Music and Audio Research Lab, New York University, New York, NY, USA}

\usepackage{lipsum}

\newcommand\blfootnote[1]{%
  \begingroup
  \renewcommand\thefootnote{}\footnote{#1}%
  \addtocounter{footnote}{-1}%
  \endgroup
}

\begin{document}

\maketitle
\begin{abstract}
We introduce a new multidimensional representation,
named \emph{eigenprogression transform}, that characterizes some essential patterns of Western tonal harmony while being equivariant to time shifts and pitch transpositions.
This representation is deep, multiscale, and convolutional in the piano-roll domain, yet incurs no prior training, and is thus suited to both supervised and unsupervised MIR tasks.
The eigenprogression transform combines ideas from the spiral scattering transform, spectral graph theory, and wavelet shrinkage denoising.
We report state-of-the-art results on a task of supervised composer recognition (Haydn \vs{} Mozart) from polyphonic music pieces in MIDI format.
\blfootnote{This work is supported by the ERC InvariantClass grant. The source code to reproduce experiments is released under MIT license at:
\url{www.github.com/lostanlen/ismir2018-lbd}.
The author thanks Moreno Andreatta, Joanna Devaney, Peter van Kranenburg, St\'{e}phane Mallat, Brian McFee, and Gissel Velarde for helpful comments.}
\end{abstract}

\section{Eigentriads}\label{sec:eigentriads}
Let $\boldsymbol{x}[t,p]\in\mathcal{M}_{T,P}(\mathbb{R})$ the piano-roll matrix of a musical piece, either obtained by parsing symbolic data or by extracting a melody salience representation from audio \cite{bittner2017ismir}.
The constant $T$ (\resp{} $P$) is typically equal to $2^{10}$ (\resp{} $2^7$).
Within the framework of twelve-tone equal temperament, we define the major and minor triads as the tuples $\mathcal{I}_1 = (0, 4, 7)$ and $\mathcal{I}_0 =(0, 3, 7)$.
For each quality $q\in\mathbb{Z}_2$ and frequency $\beta \in \mathbb{Z}_3$, let
\begin{equation}
\boldsymbol{\psi}_{\beta_1,q}^{\mathrm{triad}}[p] = \sum_{n=1}^{3} \exp\left(2\pi \mathrm{i} \dfrac{\beta n}{3}\right) \boldsymbol{\delta}\big[p - \mathcal{I}_q[n]\big],
\end{equation}
where $\boldsymbol{\delta}[p-\mathcal{I}_q[n]]$ is the Kronecker delta symbol, equal to one if $p=\mathcal{I}_q [n]$ and zero otherwise.
Let $\mathcal{G}_q$ the induced subgraph of $\mathcal{I}_q$, where $\mathcal{I}_q$ is understood as a set of vertices in $\mathbb{Z}_P$.
Observe that $\{p \mapsto \boldsymbol{\psi}_{\beta,q}^{\mathrm{triad}}\}_\beta$ consists of the eigenfunctions of the unnormalized Laplacian
matrix of $\mathcal{G}_q$:
\begin{equation}
\mathbf{L}^{\mathrm{triad}}_q [p,p^\prime] = \vert \mathcal{I}_q \vert \boldsymbol{\delta}[p\in\mathcal{I}_q] \boldsymbol{\delta}[p^\prime \in\mathcal{I}_q] - \boldsymbol{\delta}[p \overset{\mathcal{G}_q}{\sim} p^\prime]
\end{equation}
As a result, we propose to name \emph{eigentriads} the complex-valued signals $\boldsymbol{\psi}_{\beta_1,q}^{\mathrm{triad}}$ in pitch space.
We construct a multiresolution convolutional operator in the piano-roll domain by separable interference
$\Psi_{(\alpha_1,\beta_1,q)}^{\mathrm{triad}}[t,p] = \boldsymbol{\psi}_{\alpha_1}[t] \boldsymbol{\psi}_{\beta_1,q}^{\mathrm{triad}}[p]$
between the aforementioned eigentriads and a family of temporal Gabor wavelets
\begin{equation}
\boldsymbol{\psi}_{\alpha_1}[t] =
\alpha_1 \exp\left(-\dfrac{\alpha_1^2 t^2}{2\sigma^2}\right) \exp(\mathrm{i}\alpha_1\xi t)
\end{equation}
for $t \in \llbracket 0; T \llbracket$.
We set $\xi = \frac{2\pi}{3}$, $\sigma=0.1$, $\xi=\frac{2\pi}{3}$, and $\log_2 \alpha_1 \in \llbracket 0; \log_2 T \rrbracket$.
We define the \emph{eigentriad transform} of $\boldsymbol{x}$ as the rank-five tensor resulting from the complex modulus of all convolutions between $\boldsymbol{x}$ and multivariable wavelets $\Psi_{(\alpha_1,\beta_1,q)}^{\mathrm{triad}}[t,p]$:
\begin{multline}
\mathbf{U}_1(\boldsymbol{x})[t,p,q,\alpha_1,\beta_1] =
\left \vert \boldsymbol{x} \ast \Psi_{(\alpha_1,\beta_1,q)}^{\mathrm{triad}} \right \vert [t,p] \\
= \left \vert \sum_{t^\prime = 0}^{T-1} \sum_{p^\prime = 0}^{P-1} \boldsymbol{x}\left[t^\prime,p^\prime\right] \Psi_{(\alpha_1,\beta_1,q)}^{\mathrm{triad}}\left[t-t^\prime,p-p^\prime\right]\right \vert,
\end{multline}
where the difference in $t$ (\resp{} in $p$) is computed in $\mathbb{Z}_{T}$ (\resp{} in $\mathbb{Z}_P$).
By averaging the tensor $\mathbf{U}_1 (\boldsymbol{x})$ over the dimensions of time $t$, pitch $p$, and triad quality $q$, one obtains the matrix
\begin{equation}
\mathbf{S}_1 (\boldsymbol{x})[\alpha_1, \beta_1]
=
\sum_{t\in\mathbb{Z}_T} \sum_{p\in\mathbb{Z}_P} \sum_{q\in\mathbb{Z}_2}
\mathbf{U}_1 (\boldsymbol{x})[t,p,q,\alpha_1,\beta_1]. \label{eq:S1}
\end{equation}
The operator $\boldsymbol{x} \mapsto \mathbf{S}_1 (\boldsymbol{x})$ characterizes the relative amounts of ascending triads ($\beta_1 = 1$), descending triads ($\beta_1 = -1$), and perfect chords ($\beta_1 = 0$) at various temporal scales $\alpha_1$ in the piece $\boldsymbol{x}$, while keeping a relatively low dimensionality, equal to $3\log_2 T$.
The averaging along variables $t$ and $p$ involved in Equation \ref{eq:S1} guarantees that $\mathbf{S}_1$ is invariant to the action of any temporal shift operator $\boldsymbol{\tau}_{\Delta t}:\boldsymbol{x}[t,p] \mapsto \boldsymbol{x}[t+\Delta t,p]$, as well as any pitch transposition operator $\boldsymbol{\pi}_{\Delta p}:\boldsymbol{x}[t,p] \mapsto \boldsymbol{x}[t,p+\Delta p]$:
\begin{equation}
\forall \Delta t \in \mathbb{Z_T},
\forall \Delta p \in \mathbb{Z_P},
\mathbf{S}_1 (\boldsymbol{\pi}_{\Delta p} \circ \boldsymbol{\tau}_{\Delta t} \circ \boldsymbol{x}) = \mathbf{S}_1 (\boldsymbol{x}).
\end{equation}
Furthermore, the averaging across triad qualities $q$ implies approximate invariance to tonality, in the sense that replacing major triads by minor triads and vice versa in $\boldsymbol{x}$ (insofar as this is feasible in the signal $\boldsymbol{x}$ at hand) does not affect the matrix $\mathbf{S}_1(\boldsymbol{x})$.
From the standpoint of serialist music theory \cite{babbitt1960quarterly}, the presence of signed eigentriad frequencies $\beta_1 = \pm 1$ ensures that $\mathbf{S}_1$
is not invariant to retrogradation $\boldsymbol{R}: \boldsymbol{x}[t,p] \mapsto \boldsymbol{x}[-t,p]$, \ie{} time reversal:
\begin{equation}
(\boldsymbol{R} \circ \boldsymbol{x}) \neq \boldsymbol{x} \implies
\mathbf{S}_1 (\boldsymbol{R} \circ \boldsymbol{x}) \neq \mathbf{S}_1 (\boldsymbol{x}).
\end{equation}
However, the averaging across triad qualities $q$ causes $\mathbf{S}_1$ to be invariant to inversion
$\boldsymbol{I} : \boldsymbol{x}[t,p] \mapsto \boldsymbol{x}[t,-p]$, \ie{} reversal of the pitch axis:
\begin{equation}
(\boldsymbol{I} \circ \boldsymbol{x}) \neq \boldsymbol{x} \;\;\not\!\!\!\implies
\mathbf{S}_1 (\boldsymbol{I} \circ \boldsymbol{x}) \neq \mathbf{S}_1 (\boldsymbol{x}).
\end{equation}
The above property hinders the accurate modeling of chord progressions in the context of Western tonal music.
Indeed, $\mathbf{S}_1$ fails to distinguish a perfect major cadence ($\mathsf{C}^{\mathsf{maj}}\rightarrow\mathsf{F}^{\mathsf{maj}}$) from a plagal minor cadence ($\mathsf{F}^{\mathsf{min}} \rightarrow \mathsf{C}^{\mathsf{min}}$), as one proceeds from the other by involution with $\mathbf{I}$.
More generally, the eigentriad transform may suffice for extracting the quality of isolated chords, but lacks longer-term context of harmonic tension and release so as to infer the tonal functions of such chords (\eg{} tonic \vs{} dominant).

\section{Eigenprogressions}\label{sec:eigenprogressions}
In this section, we introduce a second multidimensional feature $\mathbf{U}_2$, built on top of $\mathbf{U}_1$ and named \emph{eigenprogression transform}, that aims at integrating harmonic context in Western tonal music while still respecting the aforementioned requirements of invariance to global temporal shifts $\boldsymbol{\tau}_{\Delta t}$ and pitch transpositions $\boldsymbol{\pi}_{\Delta p}$.
We begin by defining the Tonnetz as an undirected graph over the $24$ vertices of triads $(p,q)\in\mathbb{Z}_{12} \times \mathbb{Z}_2$.
Let $\mathcal{J}_1 = 4$ (\resp{} $\mathcal{J}_0 = 3$) the number of semitones in a major (\resp{} minor) third.
The unnormalized Laplacian tensor of the Tonnetz is
\begin{eqnarray}
\mathbf{L}^{\mathrm{Tonnetz}}[p,q,q^\prime,p^\prime] &= &
\Big( \boldsymbol{\delta}[(-1)^q (p-p^\prime) \in\mathcal{J}_q] \nonumber  \\
& + & \boldsymbol{\delta}[(-1)^{q^\prime} (p-p^\prime) \in\mathcal{J}_{q^\prime}] \nonumber  \\
& + & \boldsymbol{\delta}[p-p^\prime]\Big) \times \boldsymbol{\delta}[q-q^\prime+1] \nonumber \\ 
& - & 3 \boldsymbol{\delta}[p-p^\prime]\boldsymbol{\delta}[q-q^\prime].
\end{eqnarray}
We define eigenvalues $\lambda$ of the Laplacian tensor, and corresponding eigenvectors $\mathbf{v}$, as the solutions of the following equation:
\begin{eqnarray}
(\mathbf{L}^{\mathrm{Tonnetz}} \overline{\overline{\otimes}} \mathbf{v})[p,q]
& = & \sum_{p^\prime,q^\prime} \mathbf{L}[p,q,q^\prime,p^\prime] \mathbf{v}[p^\prime,q^\prime] \nonumber \\
& = &
\lambda \mathbf{v}[p,q].
\label{eq:eigenvalues}
\end{eqnarray}
The number of distinct eigenvalues $\lambda_1 \ldots \lambda_K$ satisfying Equation \ref{eq:eigenvalues} is equal to $K$, and their associated eigensubspaces form a direct sum:
$\mathbf{V}_1 \oplus \ldots \oplus \mathbf{V}_K = \mathcal{M}_{P,2}(\mathbb{R})$.
For values of $k$ such that $\dim \mathbf{V}_k = 1$, we define the $k^\textrm{th}$ eigenprogressions as
$\boldsymbol{\psi}^{\mathrm{Tonnetz}}_k [p,q] = \mathbf{v}_k[p,q]$, where $\mathbf{v}_k \in \mathbf{V}_k$ and $\Vert \mathbf{v}_k \Vert_2 = 1$.
In contrast, for values of $k$ such tat $\dim \mathbf{V}_k = 2$, we arbitrarily select two vectors $\mathbf{v}_k^{\mathbb{R}\mathrm{e}}$ and $\mathbf{v}_k^{\mathbb{I}\mathrm{m}}$ satisfying $\mathbf{v}_k^{\mathbb{R}\mathrm{e}} \perp \mathbf{v}_k^{\mathbb{I}\mathrm{m}}$, $\Vert\mathbf{v}_k^{\mathbb{R}\mathrm{e}}\Vert_2 = \Vert\mathbf{v}_k^{\mathbb{I}\mathrm{m}}\Vert_2 = 1$, and $\mathrm{span}(\{\mathbf{v}_k^{\mathbb{R}\mathrm{e}},\mathbf{v}_k^{\mathbb{I}\mathrm{m}}\})=\mathbf{V}_k$; and define eigenprogressions as
\begin{equation}
\boldsymbol{\psi}^{\mathrm{Tonnetz}}_{\beta_2}[p,q] =
\mathbf{v}_k^{\mathbb{R}\mathrm{e}}[p,q] + \mathrm{i} \mathbf{v}_k^{\mathbb{I}\mathrm{m}}[p,q].
\end{equation}
We define multivariable eigenprogression wavelets as
\begin{equation}
\boldsymbol{\Psi}_{(\alpha_2,\beta_2,\gamma_2)}^{\mathrm{prog}}[t,p,q] = \boldsymbol{\psi}_{\alpha_2}[t] \boldsymbol{\psi}_{\beta_2}^{\mathrm{Tonnetz}}[p,q] \boldsymbol{\psi}_{\gamma_2}^{\mathrm{spiral}}\left[ p \right],
\end{equation}
where $\boldsymbol{\psi}_{\alpha_2}$ is a temporal Gabor wavelet of frequency and $\boldsymbol{\psi}_{\gamma_2}^{\mathrm{spiral}}$ is a Gabor wavelet on the Shepard pitch spiral \cite{lostanlen2015dafx}:
\begin{equation}
\boldsymbol{\psi}_{\gamma_2}^{\mathrm{spiral}}[p] = 
\gamma_2 \exp\left(-\dfrac{\gamma_2^2 \lfloor \frac{p}{12} \rfloor^2}{2\sigma^2}\right) \exp\left(\mathrm{i}\gamma_2\xi \left\lfloor \frac{p}{12} \right\rfloor\right),
\end{equation}
wherein $\gamma_2 \in \{0, \pm1\}$.
We define the \emph{eigenprogression transform} of $\boldsymbol{x}$ as the following rank-eight tensor:
\begin{eqnarray}
\mathbf{U}_2(\boldsymbol{x})[t,p,q,\alpha_1,\beta_1,\alpha_2,\beta_2,\gamma_2]
= \qquad  \qquad \qquad \; \nonumber \\
\Big\vert \mathbf{U}_1(\boldsymbol{x})\overset{t,p,q}{\ast} \mathbf{\Psi}^\mathrm{prog}_{(\alpha_2,\beta_2,\gamma_2)} \Big\vert[t,p,q].
\end{eqnarray}
At first sight, Equation \ref{eq:eigenvalues} suffers from an identifiability problem.
Indeed, a different choice of basis for $\mathbf{V}_k$ would incur a phase shift and/or a complex conjugation of the convolutional response $\mathbf{U}_1(\mathbf{x}) \overset{p,q}{\ast} \boldsymbol{\psi}^{\mathrm{Tonnetz}}_k$.
Yet, because the eigenprogression transform consists of Gabor wavelets (\ie{} with a symmetric amplitude profile) and is followed by a complex modulus operator, such differences in phase and/or spin are eventually canceled and thus have no effect on the outcome of the transform.
Consequently, we pick one arbitrary pair $(\mathbf{v}_k^{\mathbb{R}\mathrm{e}}, \mathbf{v}_k^{\mathbb{I}\mathrm{m}})$ for each subspace $\mathbf{V}_k$, without loss of generality.

\section{Experiments}
We evaluate the eigenprogression transform on a task of supervised composer recognition between Haydn and Mozart string quartets \cite{vankranenburg2005chapter}.
After averaging along time $t$ and pitch $p$, we standardize each feature in the rank-five tensor
\begin{multline}
\mathbf{S}_2 (\boldsymbol{x})[\alpha_1,\beta_1,\alpha_2,\beta_2,\gamma_2] =
\\
\sum_{t,p,q} \mathbf{U}_2 (\boldsymbol{x})[t,p,q,\alpha_1,\beta_1,\alpha_2,\beta_2,\gamma_2]
\end{multline}
to null mean and unit variance.
Then, we train a linear support vector machine with $C=10^4$, and report results with leave-one-out cross-validation.
The ablation study in Table \ref{table:results} confirm that all the five scattering variables ($\alpha_1$, $\beta_1$, $\gamma_1$, $\alpha_2$, $\beta_2$, $\gamma_2$) are beneficial to both sparsity and accuracy.
However, because the dataset string quartet movements contains only $107$ examples in total, the full eigenprogression transform (in dimension $8385$) is exposed to statistical overfitting.
For the sake of simplicity, rather than running a feature selection algorithm, we apply wavelet shrinkage denoising, \ie{} we keep $1119$ coefficients of largest energy on average, summing up to $50\%$ of the total energy.
This adaptive procedure has been proven to be near-optimal in the context of wavelet bases \cite{donoho1994biometrika}.
It leads to a state-of-the-art classification accuracy of $82.2\%$.

\begin{table}[]
\begin{tabular}{lllllrrr}
& & & & & dim. & $\ell_1/\ell_2$ & acc. (\%) \\
\cite{vankranenburg2005chapter} & & & & & & & 79.4 \\
\cite{velarde2016ismir} & & & & & & & 80.4 \\
\hline
$\alpha_1$ &  & &  & & 8 & 2.6 & 67.3 \\
$\alpha_1$ & $\beta_1$  & & & & 24 & 4.6 & 71.0 \\
$\alpha_1$ & $\beta_1$  & $\alpha_2$ & & & 129 & 6.1 & 72.0 \\
$\alpha_1$ & $\beta_1$  & $\alpha_2$ & $\beta_2$ & & 1677 & 17.0 & 76.7 \\
$\alpha_1$ & $\beta_1$  & $\alpha_2$ & $\beta_2$ & $\gamma_2$ & 8385 & 42.4 & 77.6 \\
$\alpha_1$ & $\beta_1$  & $\alpha_2$ & $\beta_2$ & $\gamma_2$ & 1119 & 22.3 & \textbf{82.2} \\
\end{tabular}
\caption{Comparison between the eigenprogression transform and other transforms of smaller tensor rank, in terms of dimensionality, sparsity ($\ell_1 / \ell_2$ ratio), and accuracy on a supervised task of composer recognition.}
\label{table:results}
\end{table}

\section{Conclusion}
We have diagonalized the Laplacian of the Tonnetz graph and derived a multivariable scattering transform, named eigenprogression transform, that captures some local harmonic context in Western tonal music.
Although the numerical example we gave was a task of composer recognition, the eigenprogression transform could, in principle, addresss other MIR tasks in the future, including cover song retrieval, key estimation, and structure analysis.

\newpage
\bibliography{ISMIRtemplate}

%
%
%
%

\end{document}